\documentclass[11pt]{article}
\usepackage{verbatim}
\usepackage{amsmath,amssymb}
\makeatletter \@addtoreset{equation}{section} \makeatother

\newcommand{\noi}{\vspace{12pt}\noindent}
\newcommand{\beq}{\begin{equation}}
\newcommand{\eeq}{\end{equation}}
\newcommand{\bea}{\begin{eqnarray}}
\newcommand{\eea}{\end{eqnarray}}

\newcommand{\e}[1]{{(\ref{#1})}}
\newcommand{\eq}[1]{{eq.\ (\ref{#1})}}
\newcommand{\es}[2]{{(\ref{#1}) and (\ref{#2})}}

\newcommand{\Ref}[1]{{Ref.~\cite{#1}}}
\newcommand{\mb}[1]{{\mbox{${#1}$}}}
\newcommand{\equi}[1]{\stackrel{{#1}}{=}}

\newcommand{\cM}{{\cal M}}

\newcommand{\cO}{{\cal O}}
\newcommand{\cU}{{\cal U}}
\newcommand{\cV}{{\cal V}}
\newcommand{\cZ}{{\cal Z}}

\newcommand{\ie}{{${ i.e., \ }$}}
\newcommand{\eg}{{${ e.g., \ }$}}

\newcommand{\cf}{{cf.\ }}

\newcommand{\aka}{{also known as }}

\newcommand{\wrtt}{{with respect to the }}

\newcommand{\rhs}{{right-hand side }}
\renewcommand{\~}{ \ }
\renewcommand{\=}{ \ = \ }

\renewcommand{\Tilde}{\widetilde}
\renewcommand{\Bar}{\overline}
\newcommand{\eps}{\varepsilon^{}}
\newcommand{\p}{\!{}^{}}
\newcommand{\q}{{}^{}}

\newcommand{\Exp}{{\rm Exp}}
\newcommand{\Ln}{{\rm Ln}}

\newcommand{\cl}{{\rm cl}}

\newcommand{\Hf}{\frac{1}{2}}

\newcommand{\Ih}{\frac{i}{\hbar}}
\newcommand{\Hi}{\frac{\hbar}{i}}

\newcommand{\twostack}[2]{\begin{array}{c} \lower.8ex\hbox{${#1}$}
                     \cr \raise.8ex\hbox{${#2}$} \end{array}}
\newcommand{\deder}[1]{\frac{ 
 \stackrel{\raise.2ex\hbox{$\leftarrow$}}{\delta^{r}}   } 
 {   \delta {#1}}  }
\newcommand{\dedel}[1]{\frac{ 
 \stackrel{\lower.3ex \hbox{$\rightarrow$}}{\delta^{\ell}}   }
 {   \delta {#1}}  }
\newcommand{\papar}[1]{\frac{  
 \stackrel{\raise.2ex\hbox{$\leftarrow$}}{\partial^{r}}   } 
 {   \partial {#1}}  }
\newcommand{\papal}[1]{\frac{ 
 \stackrel{\lower.3ex \hbox{$\rightarrow$}}{\partial^{\ell}}   }
 {   \partial {#1}}  }
\newcommand{\explpapar}[1]{\frac{  
 \stackrel{\raise.2ex\hbox{$\leftarrow$}}{\partial^{r}_{\rm expl}}   } 
 {   \partial {#1}}  }
\newcommand{\explpapal}[1]{\frac{ 
 \stackrel{\lower.3ex \hbox{$\rightarrow$}}{\partial^{\ell}_{\rm expl}}   }
 {   \partial {#1}}  }
\newcommand{\expldeder}[1]{\frac{  
 \stackrel{\raise.2ex\hbox{$\leftarrow$}}{\delta^{r}_{\rm expl}}   } 
 {   \delta {#1}}  }
\newcommand{\expldedel}[1]{\frac{ 
 \stackrel{\lower.3ex \hbox{$\rightarrow$}}{\delta^{\ell}_{\rm expl}}   }
 {   \delta {#1}}  }
\newcommand{\implpapar}[1]{\frac{  
 \stackrel{\raise.2ex\hbox{$\leftarrow$}}{\partial^{r}_{\rm impl}}   } 
 {   \partial {#1}}  }
\newcommand{\implpapal}[1]{\frac{ 
 \stackrel{\lower.3ex \hbox{$\rightarrow$}}{\partial^{\ell}_{\rm impl}}   }
 {   \partial {#1}}  }
\newcommand{\rpa}[1]{{ 
 \stackrel{\raise.2ex\hbox{$\leftarrow$}}{\partial^{r}_{#1}}   }}
\newcommand{\lpa}[1]{{ 
 \stackrel{\lower.3ex\hbox{$\rightarrow$}}{\partial^{\ell}_{#1}}  }}

\begin{document}
\thispagestyle{empty}
\title{\Large{\bf Reparametrization-Invariant Effective Action \\
in Field-Antifield Formalism}}
\author{{\sc Igor~A.~Batalin}$^{a}$ and {\sc Klaus~Bering}$^{b}$ \\~\\
$^{a}$I.E.~Tamm Theory Division\\
P.N.~Lebedev Physics Institute\\Russian Academy of Sciences\\
53 Leninsky Prospect\\Moscow 119991\\Russia\\~\\
$^{b}$Institute for Theoretical Physics \& Astrophysics\\
Masaryk University\\Kotl\'a\v{r}sk\'a 2\\CZ--611 37 Brno\\Czech Republic}
\maketitle
\vfill
\begin{abstract}
We introduce classical and quantum antifields in the
reparametrization-invariant effective action, and derive a deformed classical
master equation. 
\end{abstract}
\vfill
\begin{quote}
PACS number(s): 03.70.+k; 11.10.-z; 11.10.Ef; 11.15.-q;  
\hrule width 5.cm \vskip 2.mm \noindent 
$^{a}${\small E--mail:~{\tt batalin@lpi.ru}} \hspace{10mm}
$^{b}${\small E--mail:~{\tt bering@physics.muni.cz}} \\
\end{quote}

\newpage

\section{Introduction}
\label{secintro}

\noi
It is well-known that the basic/standard notion of effective action 
$\Gamma(\Phi)$ in quantum field theory is not reparametrization-invariant, \cf 
Sections~\ref{secstandardlegendre}--\ref{secstandardzandeffact}. 
A remedy was proposed by Vilkovisky \cite{vilk84a,vilk84b,dewitt87} by using
a connection $\Gamma^{\gamma}\q_{\alpha\beta}$ on the field configuration
manifold $\cM$, \cf Sections~\ref{seclogmap}--\ref{secrepineffact01} 
and Appendix~\ref{secmetricsynge}. In this paper, we amend the 
reparametrization-invariant construction with antifields, and
develop the corresponding field-antifield formalism \cite{bv81,bv83,bv84,bv85}.
We derive a Ward identity \e{wi01} and a deformed classical master equation
\e{cme01}, \cf Section~\ref{secrepineffact02} and Appendix~\ref{secextform}. 
The resulting approach works in principle for an arbitrary gauge theory. A 
manifest superfield approach is considered in Appendix~\ref{secsuper}.

\noi
In the following we will use DeWitt condensed notation.

\section{Legendre Transformation}
\label{secstandardlegendre}

\noi
In quantum field theory, one often performs a Legendre transformation 
\beq
W\p_{c}(J) -\Gamma(\Phi) \~\equiv\~ J\p_{\alpha}\Phi^{\alpha}  
\label{legendretransf01}
\eeq
to change variables $J\p_{\alpha}\leftrightarrow \Phi^{\alpha}$
from sources $J\p_{\alpha}$ to classical fields $\Phi^{\alpha}$. 
Here $W\p_{c}\equiv\Hi\ln\cZ$ is the generating action for connected diagrams
and $\Gamma(\Phi)$ is the effective action.
One usually takes $n$ implicit relations
\beq
J\p_{\alpha}\=J\p_{\alpha}(\Phi)
\qquad\Leftrightarrow\qquad 
\Phi^{\alpha}\=\Phi^{\alpha}(J)\label{standardimplicitrel01}
\eeq
to be of the form 
\beq
\Phi^{\alpha} \= (\papal{J\p_{\alpha}}W\p_{c})   
\qquad \stackrel{\e{legendretransf01}}{\Leftrightarrow} \qquad
J\p_{\alpha} \= -(\Gamma \papar{\Phi^{\alpha}})\~.\label{legendretransf02}
\eeq
We stress that although the relations \e{legendretransf02} are the most natural 
choice of implicit relations \e{standardimplicitrel01}, they are not the only 
possibility, as we shall see in Section~\ref{secrepineffact01}.

\section{Standard Partition Function and Effective Action}
\label{secstandardzandeffact}

\noi
The standard (non-reparametrization-invariant) partition function $\cZ(J)$ 
depends on sources $J\p_{\alpha}$
\beq
e^{\Ih W\p_{c}(J)}\~\equiv\~\cZ(J) 
\~:=\~\int \!d\mu\~ e^{\Ih (W(\varphi) + J\p_{\alpha}\varphi^{\alpha})}\~.
\label{nonrepzet01}
\eeq
The quantum average is
\beq
\langle F \rangle\q_{J}
\~:=\~\frac{1}{\cZ(J)}\int \!d\mu\~ e^{\Ih (W(\varphi) 
+ J\p_{\alpha}\varphi^{\alpha})}F\~,
\qquad F\=F(\varphi)\~.
\label{nonrepquanaverage01}
\eeq
Here $W=W(\varphi)$ is a (gauge-fixed) quantum action, and $\varphi^{\alpha}$ is
the quantum field/integration variable of Grassmann parity $\eps_{\alpha}$.
The level-zero\footnote{The multi-level formalism was introduced in
Refs.\ \cite{bt92,bt93,bt94,bms95,bt96} and reviewed in \Ref{bbd06}.} 
measure in the path integral \e{nonrepzet01} is
\beq
d\mu\~:=\~\rho [d\varphi]
\=\rho\prod_{\alpha}d\varphi^{\alpha}
\~,\qquad \rho\=\rho(\varphi)\~.\label{pimeasure01}
\eeq
The effective action is defined as
\beq
e^{\Ih\Gamma(\Phi)}\~:=\~  \left. \int d\mu\~ e^{\Ih (W(\varphi) + 
J\p_{\alpha}(\varphi^{\alpha}-\Phi^{\alpha} ))} \right|_{J=J(\Phi)} \~,
\label{nonrepeffact01}
\eeq
where the implicit relations \e{standardimplicitrel01} are given by the
standard Legendre relations \e{legendretransf02}. Standard reasoning yields 
that
\beq
\Phi^{\alpha}(J)\~\equi{\e{legendretransf02}}\~
(\papal{J\p_{\alpha}}W\p_{c})
\=(\explpapal{J\p_{\alpha}}W\p_{c})
\~\equi{\e{nonrepzet01}}\~ \langle \varphi^{\alpha}\rangle\q_{J} \~,
\label{clasquan01}
\eeq
while
\beq
(\Gamma\explpapar{\Phi^{\alpha}})
\~\equi{\e{nonrepeffact01}}\~ - J\p_{\alpha}(\Phi)\~,
\eeq
and
\beq
(\Gamma\papar{\Phi^{\alpha}}) - (\Gamma\explpapar{\Phi^{\alpha}})
\~\equiv\~(\Gamma\implpapar{\Phi^{\alpha}})
\~\equi{\e{nonrepeffact01}+\e{clasquan01}}\~0~.
\label{impldiff01}
\eeq
Moreover, 
\beq
 \langle F(\varphi)\rangle\q_{J}
\~\equi{\e{nonrepzet01}+\e{nonrepquanaverage01}}\~
e^{-\Ih W\p_{c}(J)}\~F\left(\Hi\explpapal{J\p_{\alpha}}\right)e^{\Ih W\p_{c}(J)}
\=F(\Phi(J))+\cO(\hbar) \~. \label{clasquan01fct}
\eeq
Here explicit dependence ``expl'' means dependence that is not via the 
implicit relations \e{standardimplicitrel01}. Note that in the standard 
Legendre transformation \e{legendretransf02}, total and explicit 
differentiations are the same.

\section{Logarithmic Map}
\label{seclogmap}

\noi
Let $(\cM,\nabla)$ be the $n$-dimensional field configuration manifold $\cM$
endowed with a torsion-free (tangent space) connection $\nabla$. Let $\cM$ have
local (position) coordinates $\varphi^{\alpha}$ with Grassmann parity 
$\eps(\varphi^{\alpha})=\eps_{\alpha}$, where $\alpha=1, \ldots, n$.

\noi
Let $\Phi$ be a fixed base point. Let $\cV \subseteq T\p_{\Phi}\cM$ be an 
sufficiently small open neighborhood of the zero (velocity) vector 
$0\in T\p_{\Phi}\cM$. The {\bf exponential map}
$\Exp\q_{\Phi}:\cV \subseteq T\p_{\Phi}\cM \to \cM$ takes a (velocity) vector 
$v\q_{\Phi}\in\cV$ and maps it to the unique point $\varphi\in \cM$ on the
manifold that is reached along a geodesic $\gamma:[t\q_{0},t\q_{1}] \to \cM$, 
\ie
\beq
\gamma(t\!=\!t\q_{0})\=\Phi\~, \qquad  
(t\q_{1}-t\q_{0})\dot{\gamma}(t\!=\!t\q_{0})\=v\q_{\Phi}\~, \qquad
\Exp\q_{\Phi}(v\q_{\Phi})\~:=\~\gamma(t\!=\!t\q_{1})\=\varphi\~.
\eeq
These formulas are invariant under affine reparametrizations $t\to a t +b $ 
of the geodesic $\gamma$. The geodesic differential equation reads
\beq
0\=(\nabla\p_{\dot{\gamma}}\dot{\gamma})^{\alpha}
\=\dot{\gamma}^{\beta}(\nabla\p_{\beta}\dot{\gamma})^{\alpha}
\=\dot{\gamma}^{\beta}\left(\partial^{\ell}_{\beta}\dot{\gamma}^{\alpha}
+\Gamma\q_{\beta}{}^{\alpha}\q_{\delta}\~\dot{\gamma}^{\delta} \right)
\=\ddot{\gamma}^{\alpha}
+(-1)^{\eps_{\beta}}\Gamma^{\alpha}\q_{\beta\delta}\~
\dot{\gamma}^{\delta}\dot{\gamma}^{\beta}\~. \label{geodesiceq01} 
\eeq
{}For a point $\varphi$ sufficiently close to the fixed point $\Phi\in\cM$ 
(technically speaking, for points in a so-called normal neighborhood 
$\varphi\in\cU(\Phi)\subseteq\cM$), there exists a unique geodesic 
$\gamma:[t\q_{0},t\q_{1}] \to \cM$ that goes from $\Phi$ to $\varphi$. 
One defines the {\bf logarithmic map} $\Ln\q_{\Phi}:\cU(\Phi) \to T\p_{\Phi}\cM$
as the inverse of the exponential map, \ie it has the corresponding initial 
velocity vector as output,
\beq
\gamma(t\!=\!t\q_{0})\=\Phi\~, \qquad \gamma(t\!=\!t\q_{1})\=\varphi\~,  \qquad
\Ln\q_{\Phi}(\varphi)\~:=\~ (t\q_{1}-t\q_{0})\~\dot{\gamma}(t\!=\!t\q_{0})\~.
\label{ln01}
\eeq
Often in the literature, the logarithmic map (\ie the initial velocity vector) 
is denoted as 
\beq
\Ln\q_{\Phi}(\varphi)
\= -\sigma^{\alpha}(\Phi,\varphi)\~ \papal{\Phi^{\alpha}}\~,\qquad  
\sigma^{\alpha}(\Phi,\varphi)
\=(t\q_{0}-t\q_{1})\~\dot{\gamma}^{\alpha}(t\!=\!t\q_{0})\~.
\label{ln02}
\eeq
The coordinate functions $-\sigma^{\alpha}(\Phi,\varphi)$ are \aka the
{\bf Riemann normal coordinates} based at $\Phi$. 
Note that the bi-local coordinate function $\sigma^{\alpha}(\Phi,\varphi)$ 
behaves geometrically as a vector \wrtt point $\Phi$ and as a scalar \wrtt
point $\varphi$. A short-distance expansion of the logarithmic map reads
\beq
\sigma^{\alpha}(\Phi,\varphi) \= (\Phi-\varphi)^{\alpha} 
-\frac{(-1)^{\eps_{\beta}}}{2} \Gamma^{\alpha}\q_{\beta\gamma}(\Phi)\~
(\Phi-\varphi)^{\gamma} (\Phi-\varphi)^{\beta}
+\cO \left((\Phi-\varphi)^{3}\right)\~.\label{sigmashortdistexpan} 
\eeq
If the Christoffel symbols $\Gamma^{\alpha}\q_{\beta\gamma}=0$ 
vanish identically in the neighborhood $\cU(\Phi)$, then the logarithmic map 
is simply given by
\beq
\sigma^{\alpha}(\Phi,\varphi)\=\Phi^{\alpha}-\varphi^{\alpha}
\qquad \text{if}\qquad \Gamma^{\alpha}\q_{\beta\gamma}\=0\~.\label{flatlimit01}
\eeq
The logarithmic map satisfies the differential equation
\beq
\sigma^{\beta}(\Phi,\varphi)\~
(\nabla^{(\Phi)}_{\beta}\sigma)^{\alpha}(\Phi,\varphi) 
\= \sigma^{\alpha}(\Phi,\varphi)\~. \label{ln03}
\eeq

\section{Reparametrization-Invariant Effective Action}
\label{secrepineffact01}

\noi
The standard effective action \e{nonrepeffact01} is not invariant under 
reparametrizations of the quantum field $\varphi^{\alpha}$ and the classical
field $\Phi^{\alpha}$. 

\noi
The source $J\p_{\alpha}$ behaves by definition as a co-vector (scalar) under 
reparametrizations of the point $\Phi$ (the point $\varphi$), respectively. 
In particular the term $J\p_{\alpha}\Phi^{\alpha}$ is {\em not} a scalar under
reparametrizations $\Phi^{\alpha}\to \Phi^{\prime \beta}\= f^{\beta}(\Phi)$.
Since we want to maintain \eq{legendretransf01}, it therefore becomes 
impossible to make both $W\p_{c}$ and $\Gamma$ reparametrization-invariant
quantities simultaneously. We will focus on the latter, \ie the effective
action $\Gamma$.

\noi
A reparametrization-invariant effective action can be achieved by using the
logarithmic map \cite{vilk84a,vilk84b}
\beq
e^{\Ih\Gamma(\Phi)}\~:=\~ \left. \int d\mu\~ e^{\Ih (W(\varphi) - 
J\p_{\alpha}\sigma^{\alpha}(\Phi,\varphi))} \right|_{J=J(\Phi)} \~.
\label{repeffact01}
\eeq
The quantum average is
\beq
\langle F \rangle
\~:=\~e^{-\Ih\Gamma(\Phi)}\left. \int \!d\mu\~ e^{\Ih (W(\varphi)
-J\p_{\alpha}\sigma^{\alpha}(\Phi,\varphi))} F\right|_{J=J(\Phi)} \~.
\label{repquanaverage01}
\eeq 
Since we assume the Legendre relation \e{legendretransf01}, the partition 
function becomes
\beq
e^{\Ih W\p_{c}(J)}\~\equiv\~\cZ(J) \~:=\~\left. \int d\mu\~ e^{\Ih (W(\varphi) 
+ J\p_{\alpha}(\Phi^{\alpha}-\sigma^{\alpha}(\Phi,\varphi))} 
\right|_{\Phi=\Phi(J)} \~.
\label{repzet01}
\eeq
Let us now elaborate on the status of the implicit dependence
\e{standardimplicitrel01}. 

\noi
If one uses the standard Legendre relations \e{legendretransf02}, one gets 
\beq
J\p_{\alpha}(\Phi) \~\equi{\e{legendretransf02}}\~ 
-(\Gamma\papar{\Phi^{\alpha}}) 
\~\equi{\e{repeffact01}}\~
\langle \left(J\p_{\beta}(\Phi)\sigma^{\beta}(\Phi,\varphi) \right)
\papar{\Phi^{\alpha}} \rangle\~, \qquad \text{(Not in use!)}
\label{legendretransf02a}
\eeq
or equivalently,
\beq
0\~\equi{\e{legendretransf02}}\~ \Phi^{\alpha}(J)-(\papal{J\p_{\alpha}}W\p_{c})
\~\equi{\e{repzet01}}\~
\langle\papal{J\p_{\alpha}} \left(J\p_{\beta}\sigma^{\beta}(\Phi(J),\varphi) 
\right) \rangle\~,  \qquad \text{(Not in use!)}
\label{legendretransf02b}
\eeq
However, we shall here not use the standard Legendre relations 
\es{legendretransf02a}{legendretransf02b}. 

\noi
Instead we shall impose $n$ reparametrization-invariant implicit conditions
\beq
(\explpapal{J\p_{\alpha}}W\p_{c}) \= \Phi^{\alpha}(J)
\qquad\Leftrightarrow\qquad 
-(\explpapal{J\p_{\alpha}}\Gamma)\~\equiv\~ 
\langle \sigma^{\alpha}(\Phi,\varphi) \rangle\=0\~, \label{averagesigmavanish01}
\eeq
as advocated by Vilkovisky \cite{vilk84a,vilk84b,dewitt87}. The main point is 
that condition \e{averagesigmavanish01} is covariant (invariant) under 
reparametrizations of the classical field $\Phi^{\alpha}$ (quantum field 
$\varphi^{\alpha}$), respectively. The condition \e{averagesigmavanish01} 
implies that the total and explicit differentiations of the effective action 
$\Gamma$ \wrtt classical field $\Phi^{\alpha}$ are the same 
\beq
(\Gamma\papar{\Phi^{\alpha}}) - (\Gamma\explpapar{\Phi^{\alpha}})
\\~\equiv\~(\Gamma\implpapar{\Phi^{\alpha}})
\~\equi{\e{repeffact01}+\e{averagesigmavanish01}}\~0~.\label{impldiff02}
\eeq
Note that the condition \e{averagesigmavanish01} means that the classical field
$\Phi^{\alpha}$ and the quantum average $\langle \varphi^{\alpha}\rangle$ may 
differ (even at the classical level). In particular, the classical
decomposition formula \e{clasquan01fct} may no longer hold.

\noi
However, if the Christoffel symbols $\Gamma^{\gamma}\q_{\alpha\beta}=0$ vanish 
identically, then 
\begin{enumerate}
\item
the reparametrization-invariant effective action \e{repeffact01} 
reduces to the standard effective action \e{nonrepeffact01}; 
\item 
the $n$ implicit conditions \e{averagesigmavanish01} reduce to the standard 
conditions
\beq
\Phi^{\alpha} \= \langle \varphi^{\alpha}\rangle
\qquad \text{if}\qquad \Gamma^{\gamma}\q_{\alpha\beta}\=0\~, 
\eeq
\cf \eq{flatlimit01}.
\end{enumerate}

\noi
{}Finally, let us mention that one could in principle perform a change 
of integration variables
\beq
\varphi^{\alpha}\~ \longrightarrow 
\varphi^{\prime \alpha}\~:=\~\sigma^{\alpha}(\Phi,\varphi)\label{pointtransf01}
\eeq
to bring the the path integral \e{repeffact01} back to the form 
\e{nonrepeffact01} (and similarly bring the average \e{averagesigmavanish01} 
back to \eq{clasquan01}), with the caveat that the new action 
$W^{\prime}(\Phi,\varphi)=W(\Phi,\sigma(\Phi,\varphi))$ and measure 
$\rho^{\prime}(\Phi,\varphi)$ would depend on the classical fields $\Phi$.

\section{Antifields}
\label{secrepineffact02}

\noi
Next we introduce quantum antifields $\varphi^{*}_{\alpha}$ and classical 
antifields $\Phi^{*}_{\alpha}$ with opposite Grassmann parity
$\eps_{\alpha}\!+\!1$ of the corresponding field variables $\varphi^{\alpha}$
and $\Phi^{\alpha}$, which in turn carry Grassmann parity $\eps_{\alpha}$.
The antifields are co-vectors under reparametrizations of
$\varphi$ and $\Phi$, respectively. The reparametrization-invariant effective
action is
\beq
e^{\Ih\Gamma(\Phi,\Phi^{*})}\~:=\~ 
\left. \int d\mu\~ e^{\Ih (W(\varphi,\varphi^{*}) 
-J\p_{\alpha}\sigma^{\alpha}(\Phi,\varphi))} 
\right|_{J=J(\Phi,\Phi^{*})}^{
\varphi^{*}=\frac{\partial \Tilde{\Psi}}{\partial \varphi}} \~.
\label{repeffact02}
\eeq
(The vertical line notation on the \rhs of \eq{repeffact02} means that the
two formulas to the right of the vertical line should be substituted into the
path integral. By definition the substitution $J=J(\Phi,\Phi^{*})$ 
counts as implicit dependence, while the substitution
$\varphi^{*}=\frac{\partial \Tilde{\Psi}}{\partial \varphi}$ counts as 
explicit dependence.) The quantum average is
\beq
\langle F \rangle
\~:=\~e^{-\Ih\Gamma(\Phi,\Phi^{*})}
\left.\int \!d\mu\~ e^{\Ih (W(\varphi,\varphi^{*})
-J\p_{\alpha}\sigma^{\alpha}(\Phi,\varphi))}F
\right|_{J=J(\Phi,\Phi^{*})}^{
\varphi^{*}=\frac{\partial \Tilde{\Psi}}{\partial \varphi}} \~.
\label{repquanaverage02}
\eeq 
The $n$ implicit relations 
$J\p_{\alpha}=J\p_{\alpha}(\Phi,\Phi^{*})$ come by definition from the $n$ 
conditions 
\beq
\langle \sigma^{\alpha}(\Phi,\varphi) \rangle\=0 \qquad \Leftrightarrow  \qquad 
J\p_{\alpha}\=J\p_{\alpha}(\Phi,\Phi^{*})  \~, 
\label{averagesigmavanish02}
\eeq
\cf condition \e{averagesigmavanish01}.
The extended gauge-fixing Fermion $\Tilde{\Psi}$ is assumed to be affine in
the classical antifields
\beq
 \Tilde{\Psi}(\Phi^{*}, \Phi, \varphi) 
\~:=\~ \Psi(\Phi, \varphi) - \Phi^{*}_{\alpha}\sigma^{\alpha}(\Phi, \varphi)\~.
\label{tildepsi01}  
\eeq
The Fermion $\Tilde{\Psi}$ is a scalar under reparametrizations of both 
$\varphi$ and $\Phi$. We stress that the antifield-free part
$\Psi=\Psi(\Phi, \varphi)$ of the gauge-fixing Fermion $\Tilde{\Psi}$ is 
allowed to depend on the classical fields $\Phi$, in contrast to the
construction in Sections~\ref{secstandardzandeffact} and \ref{secrepineffact01}.
The condition \e{averagesigmavanish02} implies that the total and explicit 
differentiations of the effective action $\Gamma$ \wrtt classical field 
$\Phi^{\alpha}$ and antifield $\Phi^{*}_{\alpha}$ are the same 
\beq
(\Gamma\papar{\Phi^{\alpha}}) - (\Gamma\explpapar{\Phi^{\alpha}})
\\~\equiv\~(\Gamma\implpapar{\Phi^{\alpha}})
\~\equi{\e{repeffact02}+\e{averagesigmavanish02}}\~0~,\qquad
(\papal{\Phi^{*}_{\alpha}}\Gamma) - (\explpapal{\Phi^{*}_{\alpha}}\Gamma)
\\~\equiv\~(\implpapal{\Phi^{*}_{\alpha}}\Gamma)
\~\equi{\e{repeffact02}+\e{averagesigmavanish02}}\~0~.
\label{impldiff03}
\eeq
The quantum master equation \cite{bv81,bv83,bv84} reads
\beq
\Delta e^{\Ih W} \=0\qquad \Leftrightarrow\qquad \Hf(W,W) 
\= i\hbar(\Delta W)\~, \label{qmew01}
\eeq
with the odd Laplacian\footnote{Here we for simplicity assume that 
the odd scalar curvature \cite{b06,b07,bb07,bb08,bb09} vanishes.}
\beq
\Delta\~:=\~\frac{(-1)^{\eps_{\alpha}}}{\rho} 
\papal{\varphi^{\alpha}} \rho  \papal{\varphi^{*}_{\alpha}}\~, 
\qquad \rho\=\rho(\varphi)\~. \label{oddlapl01}
\eeq
The Ward identities read
\beq
J\p_{\alpha}(\Phi,\Phi^{*}) (\papal{\Phi^{*}_{\alpha}}\Gamma)
\~\equi{\e{impldiff03}+\e{wia0}}\~0\~, \label{wi01}
\eeq
and
\beq
(\Gamma \papar{\Phi^{\beta}})
\~\equi{\e{impldiff03}+\e{wia1}}\~
-J\p_{\alpha}(\Phi,\Phi^{*}) C^{\alpha}\q_{\beta} \~, \qquad 
C^{\alpha}\q_{\beta}\~:=\~ 
- \explpapal{\Phi^{*}_{\alpha}}
\langle \Tilde{\Psi}\papar{\Phi^{\beta}} \rangle\~,
\label{wi02}
\eeq
see Appendix~\ref{secextform}. The Zinn-Justin/classical master equation 
becomes
\beq
\frac{1}{2}(\Gamma,\Gamma)\q_{\cl} \~\equi{\e{antibracket01}}\~
(\Gamma \papar{\Phi^{\alpha}})(C^{-1})^{\alpha}\q_{\beta} 
(\papal{\Phi^{*}_{\beta}}\Gamma)
\~\equi{\e{wi01}+\e{wi02}}\~0\~,\label{cme01}
\eeq
where we have defined a deformed antibracket of classical variables as
\beq
(f,g)\q_{\cl}\~:=\~(f \papar{\Phi^{\alpha}})(C^{-1})^{\alpha}\q_{\beta} 
(\papal{\Phi^{*}_{\beta}}g)
-(-1)^{(\eps_{f}+1)(\eps_{g}+1)}(f\leftrightarrow g)\~. \label{antibracket01}
\eeq
The deformed classical master equation \e{cme01} is our main result. 
The antibracket \e{antibracket01} may in general violate the Jacobi
identity (even at the classical level), \cf \eq{wi2}.

\noi
{}Finally, the change of integration variables \e{pointtransf01} is now part of 
a type-2 anticanonical transformation 
\beq
\varphi^{\prime \alpha} \=(\papal{\varphi^{\prime *}_{\alpha}}\Psi\q_{2})\~, 
\qquad \varphi^{*}_{\alpha} \=(\Psi\q_{2}\papar{\varphi^{\alpha}})\~, 
\label{anticantransf01}
\eeq
with a type-2 Fermionic generator 
\beq
\Psi\q_{2}(\varphi,\varphi^{\prime *}) 
\= \varphi^{\prime *}_{\alpha}\~\sigma^{\alpha}( \Phi, \varphi )\~,
\label{fermionicgen01}
\eeq
which depends on the un-primed fields and the primed antifields. The 
anticanonical transformation respects the quantum master \eq{qmew01} as well.

\vspace{0.8cm}

\noi
{\sc Acknowledgement:}~
I.A.B.\ would like to thank M.~Lenc, R.~von Unge and the Masaryk University 
for the warm hospitality extended to him in Brno. 
K.B.\ would like to thank M.~Vasiliev and the Lebedev Physics Institute for 
warm hospitality. The work of I.A.B.\ is supported by grants RFBR 11--01--00830
and RFBR 11--02--00685. The work of K.B.\ is supported by the Grant Agency 
of the Czech Republic (GACR) under the grant P201/12/G028 and 
the U.S. Naval Academy.

\appendix

\section{Extended Formalism}
\label{secextform}

In this Appendix~\ref{secextform}, we promote (for technical rather than 
fundamental/profound reasons) the quantum fields 
$\varphi^{\alpha}$, the classical fields $\Phi^{\alpha}$ and antifields 
$\Phi^{*}_{\alpha}$ to superfields
\beq
\varphi^{\alpha}(\theta)\~:=\~\varphi^{\alpha}+\lambda^{\alpha}\theta\~,\qquad
\Phi^{\alpha}(\theta)\~:=\~\Phi^{\alpha}+\Lambda^{\alpha}\theta\~,\qquad
\Phi^{*}_{\alpha}(\theta)\~:=\~\Phi^{*}_{\alpha}-\theta J\p_{\alpha}\~,
\eeq
where $\theta$ is a Fermionic parameter. Note that the superpartners of the 
classical antifields $\Phi^{*}_{\alpha}$ are (minus) the sources $J_{\alpha}$.
Our primary aim in this Appendix~\ref{secextform} is not to create a 
superfield formalism, but merely a convenient platform to derive the
pertinent Ward identities \es{wi01}{wi02}. (A treatment from a manifest 
superfield perspective is developed in the next Appendix~\ref{secsuper}.)
Our sign convention for the Berezin integral is
\beq
 \int\!d\theta\~\theta\=1\~.
\eeq
The extended effective action 
\beq
\Bar{\Gamma}\=\Bar{\Gamma}[\Phi(\cdot); \Phi^{*}(\cdot)]
\label{exteffact02}
\eeq
depends on $4n$ variables $\Phi^{\alpha}$, $\Lambda^{\alpha}$, 
$\Phi^{*}_{\alpha}$ and $J\p_{\alpha}$. It is given as a level-one path
integral
\beq
e^{\Ih \Bar{\Gamma}}
\~:=\~\int \!d\Bar{\mu}\~e^{\Ih \Bar{A}}\~, \label{exteffact01}
\eeq
with level-one path integral measure
\beq
d\Bar{\mu}\~:=\~\rho [d\varphi][d\varphi^{*}][d\lambda]\~,
\qquad \rho\=\rho(\varphi)\~,\label{pimeasure02}
\eeq
and action
\beq
\Bar{A}\~:=\~ W + \varphi^{*}_{\alpha} \lambda^{\alpha} - Y\~, \label{action02}
\eeq
where 
\beq
W\=W(\varphi,\varphi^{*}) \label{qma02}
\eeq
is the usual quantum master action, and
\beq
Y\=Y[\Phi^{*}(\cdot),\Phi(\cdot),\varphi(\cdot)]
\eeq
is given by
\beq
Y\~:=\~ \int\!d\theta\~
\Tilde{\Psi}(\Phi^{*}(\theta),\Phi(\theta),\varphi(\theta))
\~\equi{\e{tildepsi01}}\~ J\p_{\alpha} \sigma^{\alpha}(\Phi, \varphi)  
+\Tilde{\Psi} \left(\papar{\varphi^{\alpha}}\lambda^{\alpha}
+\papar{\Phi^{\alpha}}\Lambda^{\alpha}\right)\~.\label{yact}
\eeq
Later in \eq{averagesigmavanish03} we will introduce $n$ implicit relations 
$J\p_{\alpha}=J\p_{\alpha}(\Phi,\Phi^{*},\Lambda)$. In anticipation of this, we will
already now begin to distinguish between total and explicit derivative.
Note \eg that
\beq 
(\Tilde{\Psi}\papar{\Phi^{\alpha}}) \~\equi{\e{tildepsi01}}\~ 
(\Tilde{\Psi}\explpapar{\Phi^{\alpha}}) \~, \qquad
(\papal{\Phi^{*}_{\alpha}}\Tilde{\Psi}) \~\equi{\e{tildepsi01}}\~ 
(\explpapal{\Phi^{*}_{\alpha}}\Tilde{\Psi}) \~,
\eeq
as $\Tilde{\Psi}=\Tilde{\Psi}(\Phi^{*}, \Phi, \varphi)$ does not depend on $J$.

\noi
Extended Ward identity for $Y$: 
\beq
(J\p_{\alpha}\explpapal{\Phi^{*}_{\alpha}} Y) 
+Y\left(\papar{\varphi^{\alpha}}\lambda^{\alpha}
+\explpapar{\Phi^{\alpha}}\Lambda^{\alpha}\right)\=0\~.\label{extwardid01}
\eeq
The extended Ward identity \e{extwardid01} can be seen by shifting 
integration variable $\theta \to \theta +\theta_{0}$ in the formula \e{yact} for
$Y$, and collecting terms proportional to the Fermionic constant $\theta_{0}$.

\noi
Extended Ward identity for $\Bar{\Gamma}$: 
\beq
J\p_{\alpha} (\explpapal{\Phi^{*}_{\alpha}}\Bar{\Gamma}) 
+ (\Bar{\Gamma}\explpapar{\Phi^{\alpha}})\Lambda^{\alpha}\=0\~.
\label{extwardid02}
\eeq
{\sc Proof of \eq{extwardid02}}:
\bea
0&=& \int\![d\varphi][d\varphi^{*}][d\lambda] 
(-1)^{\eps_{\alpha}} \papal{\varphi^{\alpha}}
\left(\rho (\papal{\varphi^{*}_{\alpha}} e^{\Ih W}) 
e^{\Ih(\varphi^{*}_{\alpha} \lambda^{\alpha}-Y)} \right) \cr
&\equi{\e{qmew01}}& -\int\! d\Bar{\mu} \~e^{\Ih W}
(\papal{\varphi^{*}_{\alpha}}  
e^{\Ih(\varphi^{*}_{\alpha} \lambda^{\alpha}-Y)} \papar{\varphi^{\alpha}})\cr
&=& -\int\! d\Bar{\mu} \~e^{\Ih (W+\varphi^{*}_{\alpha} \lambda^{\alpha})}
(e^{-\Ih Y} \papar{\varphi^{\alpha}}\lambda^{\alpha})\cr
&\equi{\e{extwardid01}}& \int\! d\Bar{\mu} \~e^{\Ih (W+\varphi^{*}_{\alpha} 
\lambda^{\alpha})}
\left(J\p_{\alpha} \explpapal{\Phi^{*}_{\alpha}} e^{-\Ih Y}  
+ e^{-\Ih Y} \explpapar{\Phi^{\alpha}}\Lambda^{\alpha}\right)\cr
&\equi{\e{exteffact01}}&J\p_{\alpha} (\explpapal{\Phi^{*}_{\alpha}}
e^{\Ih \Bar{\Gamma}}) 
+(e^{\Ih \Bar{\Gamma}} \explpapar{\Phi^{\alpha}})\Lambda^{\alpha}\~.
\label{extwardid03}
\eea
Expansion of the extended Ward identity \e{extwardid03} around $\Lambda=0$ to 
second order in $\Lambda$:
\beq
\left.J\p_{\alpha} \explpapal{\Phi^{*}_{\alpha}}
e^{\Ih \Bar{\Gamma}}\right|_{\Lambda=0}
\=0\~, \label{wi0}
\eeq
\beq
\Ih\left. J\p_{\alpha} \explpapal{\Phi^{*}_{\alpha}} \int \!d\Bar{\mu}\~
e^{\Ih\Bar{A}} 
(\Tilde{\Psi}\explpapar{\Phi^{\beta}})\right|_{\Lambda=0} 
\= \left.e^{\Ih \Bar{\Gamma}} \explpapar{\Phi^{\beta}}\right|_{\Lambda=0}\~,
\label{wi1}
\eeq
\bea
\lefteqn{
-\Ih\left. J\p_{\alpha} \explpapal{\Phi^{*}_{\alpha}} \int \!d\Bar{\mu}\~
e^{\Ih\Bar{A}}  (\Tilde{\Psi}\explpapar{\Phi^{\beta}})
(\Tilde{\Psi}\explpapar{\Phi^{\gamma}}) \right|_{\Lambda=0} (-1)^{\eps_{\beta}}
} \cr
&=& \left. \left( \int \!d\Bar{\mu}\~e^{\Ih\Bar{A}}  
(\Tilde{\Psi}\explpapar{\Phi^{\beta}}) 
\right)\explpapar{\Phi^{\gamma}}\right|_{\Lambda=0}
-(-1)^{\eps_{\beta}\eps_{\gamma}}(\beta \leftrightarrow \gamma) \~.\label{wi2}
\eea
Extended quantum average
\beq
 \langle F \rangle \~:=\~ e^{-\Ih \Bar{\Gamma}} 
\int \!d\Bar{\mu}\~e^{\Ih\Bar{A}} F\~.\label{extrepquanaverage01}
\eeq
Expansion of the extended Ward identity \e{extwardid02} around $\Lambda=0$ to
second order in $\Lambda$:
\beq
\left.J\p_{\alpha} (\explpapal{\Phi^{*}_{\alpha}}\Bar{\Gamma})
\right|_{\Lambda=0}\=0\~,\label{wia0}
\eeq
\beq
\left. J\p_{\alpha} \explpapal{\Phi^{*}_{\alpha}} 
\langle \Tilde{\Psi}\explpapar{\Phi^{\beta}} \rangle\right|_{\Lambda=0} 
\= \left.(\Bar{\Gamma} \explpapar{\Phi^{\beta}})\right|_{\Lambda=0}\~,
\label{wia1}
\eeq
\bea
\lefteqn{
\left. J\p_{\alpha} \explpapal{\Phi^{*}_{\alpha}} \left(
\langle (\Tilde{\Psi}\explpapar{\Phi^{\beta}})
(\Tilde{\Psi}\explpapar{\Phi^{\gamma}}) \rangle
-\langle \Tilde{\Psi}\explpapar{\Phi^{\beta}} \rangle
\langle \Tilde{\Psi}\explpapar{\Phi^{\gamma}} \rangle
\right)\right|_{\Lambda=0} (-1)^{\eps_{\beta}}
} \cr
&=&i\hbar \left.\langle \Tilde{\Psi}\explpapar{\Phi^{\beta}}\rangle 
\explpapar{\Phi^{\gamma}}\right|_{\Lambda=0}
-(-1)^{\eps_{\beta}\eps_{\gamma}}(\beta \leftrightarrow \gamma) \~.\label{wia2}
\eea
The $n$ implicit constraints reads
\beq
\langle \sigma^{\alpha}(\Phi, \varphi) \rangle \= 0 
\qquad \Leftrightarrow  \qquad 
J\p_{\alpha}\=J\p_{\alpha}(\Phi,\Phi^{*},\Lambda) \~.\label{averagesigmavanish03}
\eeq
The effective action $\Gamma=\Gamma(\Phi,\Phi^{*})$ from \eq{repeffact02} 
can now be defined via the extended effective action \e{exteffact01} as
\beq
\Gamma(\Phi,\Phi^{*})\~:=\~
\Bar\Gamma(\Phi, \Lambda\!=\!0; \Phi^{*}, 
J\!=\!J(\Phi,\Phi^{*},\Lambda\!=\!0)) \~. 
\eeq

\section{Metric and Synge's World Function}
\label{secmetricsynge}

\noi
In the main text we assumed that field configuration manifold $\cM$ is
equipped with a torsionfree connection $\nabla$.
In this Appendix~\ref{secmetricsynge} we will additionally assume that
field configuration manifold $\cM$ is equipped with a (pseudo) Riemannian 
metric $g\q_{\alpha\beta}$, and that $\nabla$ is the corresponding Levi-Civita
connection. We will follow the sign conventions of \Ref{bb09}.

\subsection{Metric}
\label{secmetric}

\noi
Let there be given a (pseudo) Riemannian metric in field configuration
manifold $\cM$, \ie a covariant symmetric $(0,2)$ tensor field
\beq
ds^2 \= d\varphi^{\alpha}\~ g\q_{\alpha\beta}\~ \vee d\varphi^{\beta}\~,
\label{defg}
\eeq
of Grassmann--parity $\eps(g\q_{\alpha\beta})=\eps_{\alpha}+\eps_{\beta}$,
and of symmetry 
\beq
g\q_{\beta\alpha}
\=-(-1)^{(\eps_{\alpha}+1)(\eps_{\beta}+1)}g\q_{\alpha\beta}\~. \label{symg}
\eeq
The symmetry \e{symg} becomes more transparent if one reorders the Riemannian 
metric as
\beq
ds^{2} \= d\varphi^{\beta} \vee d\varphi^{\alpha}\~ \Tilde{g}\q_{\alpha\beta} \~,
\label{defgtilde1}
\eeq
where\footnote{ Vilkovisky \cite{vilk84a,vilk84b} assumes that the field
configuration manifold $\cM$ is Bosonic. Our superconventions are related to 
those of DeWitt \cite{dewitt87} via 
$\Tilde{g}^{\rm (here)}_{\alpha\beta} (-1)^{\eps_{\beta}} 
\equiv g^{\rm (here)}_{\alpha\beta} 
\equiv (-1)^{\eps_{\alpha}}g^{\rm (DeWitt)}_{\alpha\beta}$, and 
$\Gamma^{\gamma}{}^{\rm (here)}_{\alpha\beta}
\equiv (-1)^{\eps_{\alpha}}\Gamma^{\gamma}{}^{\rm (DeWitt)}_{\alpha\beta}$.}
\beq
g\q_{\alpha\beta}  \= \Tilde{g}\q_{\alpha\beta} (-1)^{\eps_{\beta}}\~.
\label{defgtilde2}
\eeq
Then the symmetry \e{symg} simply reads
\beq
\Tilde{g}\q_{\beta\alpha} 
\= (-1)^{\eps_{\alpha}\eps_{\beta}}\Tilde{g}\q_{\alpha\beta}\~.
\label{symgtilde}
\eeq
The Riemannian metric $g\q_{\alpha\beta}$ is assumed to be non--degenerate, \ie
there exists an inverse contravariant symmetric $(2,0)$ tensor field 
$g^{\alpha\beta}$ such that 
\beq
g\q_{\alpha\beta}\~g^{\beta\gamma} \= \delta_{\alpha}^{\gamma}\~.\label{ginv}
\eeq
The inverse metric $g^{\alpha\beta}$ has Grassmann--parity 
$\eps(g^{\alpha\beta}) = \eps_{\alpha}+\eps_{\beta}$, and symmetry 
\beq
g^{\beta\alpha} 
\= (-1)^{\eps_{\alpha}\eps_{\beta}}g^{\alpha\beta}\~. \label{symginv}
\eeq

\subsection{Levi--Civita Connection}
\label{seclcconn}

\noi
The torsion tensor is just an antisymmetrization of the Christoffel symbol
$\Gamma^{\beta}{}_{\alpha\gamma}$ \wrtt lower indices,
\beq
 T^{\alpha}\q_{\beta\gamma}\~:=\~\Gamma^{\alpha}\q_{\beta\gamma}
+(-1)^{(\eps_{\beta}+1)(\eps_{\gamma}+1)}(\beta \leftrightarrow \gamma)\~.
\label{torsiongamma}
\eeq
In particular, the Christoffel symbol
\beq
\Gamma^{\alpha}\q_{\beta\gamma}\=-(-1)^{(\eps_{\beta}+1)(\eps_{\gamma}+1)}
(\beta \leftrightarrow \gamma)
\label{torsionfree}
\eeq
is symmetric \wrtt lower indices when the connection is torsionfree.
A connection \mb{\nabla} is called {\bf metric}, if it preserves the metric
\beq
0 \= (\nabla\p_{\alpha}\Tilde{g})\q_{\beta\gamma}
\= (\papal{\varphi^{\alpha}}\Tilde{g}\q_{\beta\gamma})
-\left((-1)^{\eps_{\alpha}\eps_{\beta}}\Gamma\q_{\beta\alpha\gamma}
+(-1)^{\eps_{\beta}\eps_{\gamma}}(\beta \leftrightarrow \gamma)\right)\~. 
\label{connmetric}
\eeq
Here we have lowered the Christoffel symbol with the metric
\beq
\Gamma\q_{\alpha\beta\gamma}
\~:=\~g\q_{\alpha\delta}\Gamma^{\delta}\q_{\beta\gamma}(-1)^{\eps_{\gamma}}\~.
\label{lowerriemconn}
\eeq
The metric condition \e{connmetric} reads in terms of the contravariant 
inverse metric 
\beq
0\=(\nabla\p_{\alpha}g)^{\beta\gamma}
\~\equiv\~(\papal{\varphi^{\alpha}}g^{\beta\gamma})
+\left(\Gamma\q_{\alpha}{}^{\beta}\q_{\delta}\~g^{\delta\gamma}
+(-1)^{\eps_{\beta}\eps_{\gamma}}(\beta\leftrightarrow \gamma)\right)~.
\label{upperriemconn}
\eeq
Here we have introduced a reordered Christoffel symbol
\beq
\Gamma\q_{\alpha}{}^{\beta}\q_{\gamma} 
\~:=\~(-1)^{\eps_{\alpha}\eps_{\beta}}\Gamma^{\beta}\q_{\alpha\gamma}~.
\label{reorderedgamma}
\eeq
The Levi--Civita connection is the unique connection $\nabla$ 
that is both torsionfree \mb{T\!=\!0} and metric \e{connmetric}.
The Levi--Civita formula for the lowered Christoffel symbol in terms of
derivatives of the metric reads
\beq
2 \Gamma\q_{\gamma\alpha\beta} 
\= (-1)^{\eps_{\alpha}\eps_{\gamma}}(\papal{\varphi^{\alpha}}
\Tilde{g}_{\gamma\beta})
+(-1)^{(\eps_{\alpha}+\eps_{\gamma})\eps_{\beta}}
(\papal{\varphi^{\beta}}\Tilde{g}_{\gamma\alpha})
-(\papal{\varphi^{\gamma}}\Tilde{g}_{\alpha\beta})\~. \label{lcformula}
\eeq

\subsection{Synge's World Functional}
\label{secsyngal}

\noi
Let $\gamma:[t\q_{0},t\q_{1}] \to \cM$ be a parametrized open curve in the 
field configuration manifold $\cM$. The {\bf Synge world functional}
$\Sigma[\gamma]$ is an off-shell action functional
\beq
\Sigma[\gamma]~:=~
\int_{\gamma} \! dt ~L(t)\~,
\label{syngedefal} 
\eeq
with Lagrangian $L(t)$ given by a normalized squared distance
\beq 
L(t)\~:=\~\frac{t\q_{1}- t\q_{0}}{2} \lambda(t)\~, \qquad 
\lambda(t)\~:=\~\dot{\gamma}^{\alpha}(t) \~g\q_{\alpha\beta}(\gamma(t)) 
\~\dot{\gamma}^{\beta}(t)\~, 
\label{geodesiclagr}
\eeq
The Synge world functional $\Sigma[\gamma]$ is invariant under affine 
reparametrizations $t\to a t + b$ of the curve $\gamma$. The corresponding 
(on-shell) Euler-Lagrange equation is precisely the geodesic \eq{geodesiceq01}.
The momentum $p\q_{\alpha}(t)$ reads
\beq
p\q_{\alpha}(t)\~:=\~L(t)\papar{\dot{\gamma}^{\alpha}(t)}
\= (t\q_{1} - t\q_{0})\~\dot{\gamma}^{\beta}(t)\~g\q_{\beta\alpha}(\gamma(t))\~.
\label{mom01}
\eeq
Since there is no explicit $t$-dependence, the corresponding energy function 
\beq
h(t)\~:=\~ p\q_{\alpha}(t)\~\dot{\gamma}^{\alpha}(t) - L(t) \= L(t) \label{energy01}
\eeq
does not depend on time $t$ on-shell, \cf Noether's theorem. In particular,
the Lagrangian $L(t)$ can be pulled outside the action integral 
$\Sigma[\gamma]\approx (t\q_{1}- t\q_{0}) L(t)$ on-shell. (Here the $\approx$ 
symbol means equality modulo the geodesic \eq{geodesiceq01}. We do not use the 
$\approx$ symbol in the main text.)

\subsection{Synge's World Function}
\label{secsynge}

\noi
Let there be given two points $\Phi,\varphi\in\cM$ that are linked by a unique 
geodesic $\gamma:[t\q_{0},t\q_{1}] \to \cM$, so that 
\beq
\gamma(t\!=\!t\q_{0})\=\Phi\~, \qquad \gamma(t\!=\!t\q_{1})\=\varphi\~.
\eeq
The {\bf Synge world function} 
\beq
\sigma(\Phi,\varphi)\~:=\~ \Sigma[\gamma] 
\eeq
between the two points $\Phi$ and $\varphi$ is defined \cite{poisson04} as 
the value of the Synge world functional
$\Sigma[\gamma]$ along the geodesic $\gamma$. In other words, the Synge world 
function $\sigma$ is the associated on-shell action function for the off-shell
action functional $\Sigma$. It follows that the Synge world function 
$\sigma(\Phi,\varphi)=\sigma(\varphi,\Phi)$ is numerically precisely 
{\em half the square of the geodesic distance from $\Phi$ to
$\varphi$},
\beq
 \Hf \left[ {\rm dist}(\Phi,\varphi) \right]^{2}
\~\approx\~\Hf \left[ \int_{t\q_{0}}^{t\q_{1}} \! dt \sqrt{\lambda(t)} \right]^{2}
\~\approx\~\frac{(t\q_{1}- t\q_{0})^2}{2}\lambda(t)
\~\approx\~\Sigma[\gamma] \~\approx\~\sigma(\Phi,\varphi)\~. 
\label{syngedef02} 
\eeq
Here in \eq{syngedef02} it is implicitely understood that the curve $\gamma$
is the geodesic between $\Phi$ to $\varphi$.

\subsection{Logarithmic Map}
\label{applogmap}

\noi
The first variation of the Synge world function is determined by the end-point 
momentas
\beq
 \delta \sigma(\Phi,\varphi)
\~\approx\~p\q_{\alpha}(t\!=\!t\q_{1})\~\delta \varphi^{\alpha}-
p\q_{\alpha}(t\!=\!t\q_{0})\~\delta \Phi^{\alpha}\~, \label{firstvar01}
\eeq
so that
\beq
 \sigma\q_{\alpha}(\Phi,\varphi)
\~:=\~\sigma(\Phi,\varphi)\papar{\Phi^{\alpha}}
\~\approx\~-p\q_{\alpha}(t\!=\!t\q_{0})\
\=(t\q_{0} - t\q_{1})\~\dot{\gamma}^{\beta}(t\!=\!t\q_{0})\~g\q_{\beta\alpha}(\Phi)
\~.\label{firstvar02}
\eeq
We next raise the index of \eq{firstvar02} with the metric
\beq
 \sigma^{\alpha}(\Phi,\varphi)
\~:=\~\sigma\q_{\beta}(\Phi,\varphi)\~g^{\beta\alpha}(\Phi)
\~\approx\~(t\q_{0}-t\q_{1})\~\dot{\gamma}^{\alpha}(t\!=\!t\q_{0})
\~,\label{ln04}
\eeq
in agreement with the general definition \e{ln02} of the logarithmic map.
It follows that
\beq
  \sigma\q_{\alpha}(\Phi,\varphi)\~\sigma^{\alpha}(\Phi,\varphi)
\=2\sigma(\Phi,\varphi)\~. \label{cuteid01}
\eeq

\section{Superfield Formalism}
\label{secsuper}

\noi
In this Appendix~\ref{secsuper} we consider a manifest superfield formalism
\cite{lmr95,bbd97,bbd98} in the antisymplectic phase space 
$\{\varphi^{\alpha};\varphi^{*}_{\beta}\}$. It is natural to also promote the
quantum antifields $\varphi^{*}_{\alpha}$ to superfields 
\beq
\varphi^{*}_{\alpha}(\theta)
\~:=\~\varphi^{*}_{\alpha}-\theta \lambda^{*}_{\alpha}\~,
\eeq
with new superpartners $\lambda^{*}_{\alpha}$. In this Appendix~\ref{secsuper} 
we will not worry about manifest reparametrization-invariance in superfield 
space. The only requirement we will demand here is that the superfield
formalism in components reduces to the previous construction of
Appendix~\ref{secextform}. The superpartners $\lambda^{*}_{\alpha}$ are
immediately killed again by modifying the path integral measure 
\e{pimeasure02} to also include a delta function 
\beq
d\Bar{\mu}\~:=\~\rho [d\varphi(\cdot)][d\varphi^{*}(\cdot)]\~\delta(\lambda^{*})
\=\rho [d\varphi(\cdot)][d\varphi^{*}(\cdot)]\~
\delta\left( \int\!d\theta\~\varphi^{*}(\theta)\right)\~.\label{pimeasure03}
\eeq
The odd Laplacian \e{oddlapl01} can be written as
\beq
\Delta\~:=\~\frac{(-1)^{\eps_{\alpha}}}{\rho} \int\!d\theta\~
\dedel{\varphi^{\alpha}(\theta)} \rho \~[\frac{d}{d\theta}, 
\dedel{\varphi^{*}_{\alpha}(\theta)}]
\=\frac{1}{\rho} \int \!d\theta\~
[\frac{d}{d\theta}, \dedel{\varphi^{\alpha}(\theta)}] 
\rho\~ \dedel{\varphi^{*}_{\alpha}(\theta)}\~. \label{oddlapl02}
\eeq
Here the functional derivatives of the quantum superfields read in components
\beq
\dedel{\varphi^{\alpha}(\theta)}\=\theta \papal{\varphi^{\alpha}}
-\papal{\lambda^{\alpha}}\~, \qquad 
\dedel{\varphi^{*}_{\alpha}(\theta)}\=\theta \papal{\varphi^{*}_{\alpha}}
+(-1)^{\eps_{\alpha}}\papal{\lambda^{*}_{\alpha}}\~,
\eeq
see also Appendix B in \Ref{bbd97}. To obtain a manifest superfield 
formulation, the $\varphi^{*}_{\alpha} \lambda^{\alpha}$-term in the
$\Bar{A}$-action \e{action02} should be replaced  
\beq
\varphi^{*}_{\alpha} \lambda^{\alpha}\quad \longrightarrow \quad 
\int\!d\theta\~\varphi^{*}_{\alpha}(\theta) \varphi^{\alpha}(\theta)
\=\varphi^{*}_{\alpha} \lambda^{\alpha}-\lambda^{*}_{\alpha}\varphi^{\alpha}
\~\approx\~\varphi^{*}_{\alpha} \lambda^{\alpha}\~,
\label{action03}
\eeq
which effectively is the same as before, due to the presence of the delta 
function $\delta(\lambda^{*})$ in the path integral measure \e{pimeasure03}.

\noi
Similarly, the quantum master action and density, 
\beq
W\=W(\varphi,\varphi^{*})\quad\text{and}\quad \rho\=\rho(\varphi)\~,
\label{qma03}
\eeq
should strictly speaking be promoted to functionals of superfields,
\beq
W\=W(\varphi(\cdot),\varphi^{*}(\cdot))\quad\text{and}\quad
 \rho\=\rho(\varphi(\cdot))\~,\label{qma04}
\eeq
respectively. However in practice, this would jeopardize the r\^ole of the
$\lambda^{\alpha}$'s as Lagrange multipliers for the gauge-fixing of the 
antifields $\varphi^{*}$. 

\noi
If one adds the action term \e{action03} to the quantum master action as
\beq
\Bar{W}
\~:=\~W+\int\!d\theta\~\varphi^{*}_{\alpha}(\theta)\varphi^{\alpha}(\theta)\~,
\eeq
if one introduces an odd vector field 
\beq
V\~:=\~\int [\frac{d}{d\theta}, \varphi^{\alpha}(\theta)]\~
d\theta \~\dedel{\varphi^{\alpha}(\theta)} 
-\int [\frac{d}{d\theta}, \varphi^{*}_{\alpha}(\theta)]\~
d\theta \~\dedel{\varphi^{*}_{\alpha}(\theta)}
\=\lambda^{*}_{\alpha}\papal{\varphi^{*}_{\alpha}}
-(-1)^{\eps_{\alpha}} \lambda^{\alpha}\papal{\varphi^{\alpha}}  \label{vvf}\~,
\eeq
and if one introduces an odd scalar
\beq
\nu\~:=\~-\int\!d\theta\~[\frac{d}{d\theta}, \varphi^{\alpha}(\theta)]
\varphi^{*}_{\alpha}(\theta)
\= \int\!d\theta\~[\frac{d}{d\theta},\varphi^{*}_{\alpha}(\theta)]
\varphi^{\alpha}(\theta)
\= \lambda^{*}_{\alpha}\lambda^{\alpha}\~\approx\~0\~,\label{nu02}
\eeq
then the quantum master equation \e{qmew01} becomes
\beq
\left(\Delta + \Ih\frac{1}{\rho} V\rho + \frac{\nu}{\hbar^{2}}\right) 
e^{\Ih \Bar{W}} \=0 \qquad \Leftrightarrow\qquad
\Hf(\Bar{W},\Bar{W})  +V[\Bar{W}]- \nu \= i\hbar(\Delta \Bar{W}+V[\ln\rho])\~,
\label{qmew04}
\eeq
because
\beq
\Hf(\Bar{W},\Bar{W}) \=\Hf(W,W)-V[W]-\nu\~, \quad
V[\Bar{W}]\=V[W]+2\nu\~, \quad (\Delta \Bar{W})\=(\Delta W)-V[\ln\rho]\~.
\eeq
{}Finally let us mention, that if one introduces an odd vector field 
\beq
U\q_{\rm expl}\~:=\~-\int [\frac{d}{d\theta}, \Phi^{\alpha}(\theta)]\~
d\theta \~\expldedel{\Phi^{\alpha}(\theta)} 
-\int [\frac{d}{d\theta}, \Phi^{*}_{\alpha}(\theta)]\~
d\theta \~\expldedel{\Phi^{*}_{\alpha}(\theta)}
\=J\p_{\alpha}\explpapal{\Phi^{*}_{\alpha}}
+(-1)^{\eps_{\alpha}} \Lambda^{\alpha}\explpapal{\Phi^{\alpha}} \label{uvf} \~,
\eeq
then the extended Ward identity \e{extwardid02} becomes
\beq
U\q_{\rm expl}[\Bar{\Gamma}]\=0\~.
\label{susyextwardid02}
\eeq

\end{document}